# A Multisite Super Learner for Survival Prediction with Limited Target-Site Data

Henry Bayly[1], Yorghos Tripodis[1], Haolin Li[1,*], and Prasad Patil[1,*]

[1]Department of Biostatistics, Boston University School of Public Health, Boston, MA, USA

[*]Co-corresponding authors: haolinli@bu.edu and patil@bu.edu

**Abstract**: Survival prediction models developed in multi-site studies may transport poorly to a new site because of between-site heterogeneity in covariate effects and baseline hazards. Pooled models may fail to capture site-specific risk patterns, while site-specific models can be unstable in small cohorts. We propose a Multi-Site Super Learner (MSSL), a domain-adaptive ensemble for survival prediction that combines diverse pooled and site-specific base learners. A meta-learner optimizes the ensemble weights using source data alone or, if available, a small target-site calibration sample. In simulations, when even a small target-site sample was available, MSSL substantially reduced prediction error relative to a pooled super learner under moderate to large between-site heterogeneity and generally matched or improved upon models fit to the target-site data alone. It did so by downweighting dissimilar or noisy models and upweighting models from structurally similar sites. Without target-site data, MSSL did not improve upon the pooled super learner. In an application to predicting dementia progression using the National Alzheimer's Coordinating Center (NACC) Uniform Data Set across 10 sites, MSSL matched or outperformed standard super learning using pooled data in 9 of 10 sites. MSSL offers a practical framework for adapting multi-site survival prediction models to a new site when a small calibration sample is available.

## Introduction

Large multi-site cohort studies provide rich time-to-event (TTE) data needed to develop prediction models of disease progression.[1,2] These datasets are valuable because of their large sample sizes, primarily achieved through aggregating information across clinical centers. While pooling data across recruitment sites increases a model's ability to learn global signal, it presents challenges to model generalization. Specifically, models developed on pooled multi-site data often exhibit performance degradation when researchers attempt to apply them in new settings.[3]

A central goal of multi-site prediction is to construct a model that can generalize to new sites despite potential differences in the study populations and data collection processes. Because large cohort studies frequently recruit participants across geographic regions and clinical centers, differences in demographic composition[4], outcome prevalence[5,6], measurement protocols[7,8], and environmental exposures[9] can result in systematic differences between sites.[10] Such heterogeneity can produce differences in predictor distributions (covariate shift) or predictor-outcome relationships (concept shift) across sites, reducing the transportability of prediction models.[11] In TTE settings specifically, varying baseline hazards between clinical centers introduces further uncertainty in developing generalizable prediction models.[12] As a result, a model that performs well on average across sites may perform poorly at individual sites, particularly sites that differ substantially from the overall study population.[3,13,14]

These challenges are particularly salient in Alzheimer's Disease (AD) and dementia prediction research. Despite a vast body of literature dedicated to the development of dementia prediction models[15–19], the explicit incorporation of between-site heterogeneity has been relatively underexplored. Prediction models for AD and dementia are often developed using data from established Alzheimer's Disease Research Centers (ADRCs) like the Alzheimer's Disease

Neuroimaging Initiative (ADNI) which aggregate data across many diverse recruitment settings.[20] ADNI recruits data from approximately 60 distinct sites across North America.[21] Previous work has shown there to be significant heterogeneity in patient trajectories due to varying patient demographics, visit patterns, and diagnostic trajectories across sites.[22] While these datasets provide increasingly large sample sizes, their inherent multi-site structure introduces heterogeneity that needs to be accounted for in prediction models.[23]

Several approaches have been proposed for prediction with heterogeneous multi-site data. Pooled models aggregate data across all sites to maximize sample size but can obscure site-specific signal.[10,24] Site-specific models can capture site-specific relationships but may result in unstable predictions due to smaller sample sizes and may generalize poorly to other sites.[25,26] Ensemble learning is one potential solution because, rather than relying on a single statistical model, ensemble methods combine predictions from multiple candidate models (for example site-specific models) to construct an optimized prediction rule.[27] Super Learning (SL) is an ensemble framework for survival prediction specifically that uses cross-validation to determine the optimal combination of candidate learners without assuming a single hazard function.[28,29] While SL provides a flexible framework for survival prediction, conventional SL approaches typically do not explicitly account for between-site heterogeneity.[28,29]

An important consideration in multi-site prediction is whether there is any information on the target site or if it is completely unseen.[30] While predicting at a completely unseen site requires pure out-of-domain generalization, even a small amount of data from the target site may allow a model to borrow more heavily from source sites that are similar to the target site. This property is particularly desirable because previous work has shown that substantial between-site heterogeneity can reduce predictive performance when models are applied to completely held-out sites.[31,32]

To leverage this opportunity, we propose a Multi-Site Super Learner (MSSL) for survival prediction which extends previous work on an SL-based approach for survival prediction[28] by accounting for between-site heterogeneity. The proposed MSSL incorporates candidate learners trained using both pooled and site-specific data. When target-site data is available, the prediction rule is tuned using the information from that site, thereby allowing the model to adapt to the new domain it is predicting in. We compare the performance of the MSSL to a SL using pooled data and site-specific approaches through simulation studies under varying levels of between-site heterogeneity. Finally, we apply the MSSL to data from the National Alzheimer's Coordinating Center (NACC)[33] to predict the time to progression from mild cognitive impairment (MCI) to dementia. This application provides an opportunity to investigate whether the predictors of dementia progression are consistent in their effect across ADRCs or whether meaningful site-specific effects can be learned.

**Proposed Method**

We consider a framework consisting of $S$ distinct source sites, indexed by $s = 1, \dots, S$ and a target site $S_{target}$. Within each site, the full cohort consists of independent and identically distributed observations denoted as $\left(\mathrm{T}_{\mathrm{i}}^{(\mathrm{s})}, \Delta_{\mathrm{i}}^{(\mathrm{s})}, \mathrm{X}_{\mathrm{i}}^{(\mathrm{s})}\right)$ for individuals $\mathrm{i} = 1, \dots, \mathrm{N}_{\mathrm{s}}$. Here, $\mathrm{T}_{\mathrm{i}}^{(\mathrm{s})}$ represents the observed time for the $i^{\mathrm{th}}$ subject at site $s$, defined as the minimum of the true event time ($\mathrm{T}_{\mathrm{i},0}^{(\mathrm{s})}$) and the censoring time ($\mathrm{C}_{\mathrm{i}}^{(\mathrm{s})}$), such that $\mathrm{T}_{\mathrm{i}}^{(\mathrm{s})} = \min\left(\mathrm{T}_{\mathrm{i},0}^{(\mathrm{s})}, \mathrm{C}_{\mathrm{i}}^{(\mathrm{s})}\right)$. The event indicator, $\Delta_{\mathrm{i}}^{(\mathrm{s})} = \mathbb{I}\left(\mathrm{T}_{\mathrm{i},0}^{(\mathrm{s})} \leq \mathrm{C}_{\mathrm{i}}^{(\mathrm{s})}\right)$, takes a value of 1 if the event of interest was observed and 0 if the observation was right-censored. Finally, $X_i^{(s)} \in \mathbb{R}^p$ denotes a $p$-dimensional vector of baseline covariates, which are assumed to be measured at all sites.

For a fixed prediction horizon $\mathrm{t_0} > 0$, the binary survival outcome is defined as $\mathrm{Y_i^{(s)}(t_0)} = \mathbb{I}\left(\mathrm{T_{i,0}^{(s)}} > \mathrm{t_0}\right)$. Using this outcome, we define an adjusted risk for any proper loss function $\ell(\cdot)$ and survival function predictor $f(X)$ as the expectation of the weighted loss:

$$\mathcal{L}(\mathrm{f}) = \mathbb{E}\left[\mathrm{d_i^{(s)} c_i^{(s)}(t_0)}\ell\left(\mathrm{Y_i^{(s)}(t_0), f\left(X_i^{(s)}\right)}\right)\right]$$

where $\mathrm{c_i^{(s)}}$represents the inverse probability of censoring weights (IPCW), which corrects for the bias introduced by right-censoring at time $t_0$. Specifically,

$$c_i^{(s)}(t_0) = \frac{\mathbb{I}\left(T_i^{(s)} \leq t_0, \Delta_i^{(s)} = 1\right)}{\hat{G}^{(s)}\left(T_i^{(s)-} \mid X_i^{(s)}\right)} + \frac{\mathbb{I}\left(T_i^{(s)} > t_0\right)}{\hat{G}^{(s)}\left(t_0 \mid X_i^{(s)}\right)}$$

where $\hat{G}^{(s)}\left(t \mid X_i^{(s)}\right) = P\left(C_i^{(s)} > t \mid X_i^{(s)}\right)$ denotes an estimate of the censoring probability distribution. Additionally, the sampling weight, $\mathrm{d_i^{(s)}}$, accounts for potentially unequal inclusion probabilities under certain epidemiological study designs. For simple random sampling with equal inclusion probabilities, the weights can be set to 1.

To implement the Multi-Site Super Learner (MSSL), we employ a two-stage hierarchical ensemble approach. In the first stage, we define a candidate learner library $\mathcal{F} = \{\mathrm{f_1, f_2, \dots, f}_l\}$ consisting of $l$ distinct algorithms. For each $f_k \in \mathcal{F}$, candidate learners are trained for (1) each site individually and (2) a pooled dataset combining all sites. Combining site-specific with pooled learners enables MSSL to balance local and global signals. When target-site data is available, target-site models are included in the ensemble. This yields a total candidate library containing $\mathcal{M} = \{1, \dots, M\}$ candidate base learners. To obtain predictions for the Stage 2 meta-learning while reducing overfitting, we use $V$-fold cross-validation of the available training data $\mathcal{D}_{\text{train}}$, which consists of the $S$ source sites. Each observation $i$ is assigned to a fold $v$. For each fold $v$, the $M$ candidate models are trained using the remaining $V-1$ folds ($\mathcal{D}_{\text{train}} \setminus$ fold $v$). Specifically, site-specific base learners, including the target-site model if data is available, for site $s$ are trained using only the subset of observations from site $s$ within $\mathcal{D}_{\text{train}} \setminus$ fold $v$, whereas the pooled learner is trained using all observations outside fold v. This yields an out-of-fold prediction for each observation $i$, $\hat{S}_m^{(-v(i))}(t_0 \mid X_i)$ , where $-v(i)$ represents the data coming from fold $v$.

In the second stage, we construct a global metadata matrix $\mathbf{Z}$ whose columns correspond to the $M$ base learners and rows correspond to individuals in the meta-learning set $\mathcal{I}_{\text{meta}}$ (defined below). $\mathbf{Z}$ itself depends on the available target-site sample size $N_{\text{target}}$:

1. Zero-Shot ($N_{\text{target}} = 0$): $Z$ contains $\sum_{s=1}^{S} N_s$ rows. Thus, $\mathcal{I}_{\text{meta}} = \{(s,i): s = 1, \dots, S,\ i = 1, \dots, N_s\}$, with the columns being each learner's out-of-fold predictions for all source sites.
2. Few-Shot ($N_{\text{target}} > 0$): $\mathbf{Z}$ contains $N_{\text{target}}$ rows with $\mathcal{I}_{\text{meta}} = \{i: \text{site}_i = S_{\text{target}}\}$. Columns for target-site and pooled learners contain out-of-fold predictions. Columns for source-site learners consist of out-of-sample predictions on the available target site data (because source-site models were fit without target-site data).

Using $\mathbf{Z}$ and $Y_i(\mathrm{t}_0)$, we estimate non-negative ensemble weights $\hat{\boldsymbol{\alpha}} = (\hat{\alpha}_1, \dots, \hat{\alpha}_M)^T$. We solve for the optimal coefficients to combine the $M$ models using non-negative least squares and normalize them to sum to one:

$$\hat{\boldsymbol{\theta}} = \arg\min_{\boldsymbol{\theta} \geq 0} \sum_{i \in \mathcal{I}_{\text{meta}}} c_i\,(t_0) \left( Y_i(t_0) - \sum_{m=1}^{M} \theta_m\,[\mathbf{Z}]_{i,m} \right)^2, \qquad \hat{\alpha}_m = \frac{\hat{\theta}_m}{\sum_{m'=1}^{M} \hat{\theta}_{m'}}$$

The final MSSL prediction for a new observation $x$ is a convex combination of the $M$ base learners re-fitted on their full respective training datasets:

$$\hat{S}_{\text{MSSL}}(t_0 \mid x) = \sum_{m=1}^{M} \hat{\alpha}_m \hat{S}_m(t_0 \mid x)$$

A visual overview of the MSSL can be found in **Figure 1**.

**Simulation Study**

*Simulation Design*

To evaluate the MSSL, we designed a multi-site prediction simulation with site-specific concept shift. The simulation includes S = 5 distinct sites (sites 1-4 are source sites and site 5 is the target site), each containing 1,500 subjects. Survival event times ($T_{\text{i},0}$) were generated from a Weibull distribution with shape parameter $\gamma = 0.7$ and scale parameter $\lambda = 1.0$. The feature space consisted of ten zero mean normally distributed covariates with a first order autoregressive (AR(1)) correlation matrix $\Sigma$ with correlation parameter $\rho = 0.6$. Censoring times were generated using a combination of random censoring ($C_i \sim \text{Uniform}(0,0.03)$) and a fixed administrative study end time ($t_{\text{end}} = 0.0162$) such that $T_{i,observed} = \min(T_{i,0}, C_i, t_{\text{end}})$. This resulted in average censoring rates varying from 68.8% to 96.4%.

We induced site-specific concept shift by partitioning the S sites into two different groups: Group 1 ($s \in \{1,2\}$) and Group 2 ($s \in \{3,4,5\}$). Letting $\beta_G \in \mathbb{R}^{10}$ denote the global baseline coefficient vector, site specific coefficients were generated by: $\beta_s = \beta_G + u_s + \epsilon_s$ where $u_s \sim N(\mu_s, \sigma^2_{group}\mathbb{I}_{10})$ is a site-dependent random shift vector. $\mu_s$ controls the direction of the concept shift according to drift magnitude $\delta$. We use the word 'drift' here to represent the magnitude of concept shift present, though it should be noted that concept shift and concept drift are often used interchangeably in the machine learning literature. For Group 1: $\delta \geq 0\colon \mu_s =$ $(-\delta, -\delta, -\delta, -\delta, -\delta, 0,0,0,0,0)$. For Group 2: $\delta \geq 0{:}\, \mu_s = (\delta, \delta, \delta, \delta, \delta, 0,0,0,0,0)$. The residual term $\epsilon_s \sim N(0, \sigma_s^2 \mathbb{I}_{10})$ captures within site variance. We fixed $\sigma_{group} = 0.2$ and $\sigma_s = 0.1$ while evaluating 11 distinct drift values from 0.0 to 1.0 in increments of 0.1. This was designed to represent the scenario where some data collection centers are similar and others vary, but no two are exactly the same. To ensure robustness, for each distinct drift value we simulated 100 datasets to evaluate each model.

The learner library consisted of three types of models: Cox proportional hazards regression, Gamma regression, and random survival forests (RSF), representing semiparametric, parametric, and nonparametric approaches, respectively. This learner library follows the recommendation of Wey et al.[34] who showed that including learners with different modeling assumptions improves the flexibility of the ensemble. To benchmark the MSSL, we compared its performance against a standard SL (SSL) using pooled data (ignoring site specific stratifications) and a target site only model. The target site only comparison was used to quantify potential performance improvements from borrowing data from similar sites compared to developing a model based on the available target site data alone. For the SSL, the data across all $S$ sites are pooled together by concatenating the site-specific matrices into a single global dataset:

$$\mathbf{X}_{\text{pooled}} = \begin{bmatrix} \mathbf{X}_1 \\ \mathbf{X}_2 \\ \vdots \\ \mathbf{X}_S \end{bmatrix} \in \mathbb{R}^{N \times p}$$

Where $N = \sum_{s=1}^{S} n_s$. The learner library, $M$, therefore consists of the $l$ distinct base learners. Standard super learning is then applied using this dataset.[29]

To evaluate how effectively the MSSL could detect and leverage site-specific heterogeneity we designated Site 5 as $S_{\text{target}}$ and systematically varied the sample size available ($N_{\text{target}} \in 0,25,50,100,150$). The remaining data from $S_{\text{target}}$ was used for model evaluation. To ensure a fair comparison, all ensemble methods were provided access to the same total dataset ($\mathcal{D}_{\text{train}} \cup \mathcal{D}_{\text{target}}$). For the SSL, target-site observations were appended directly to the combined source dataset prior to model training. The ensemble weights for the SSL were then optimized using the target-site observations. This design allows us to separate performance gains attributable to target-site data from those attributable to the multi-site library structure.

Predictive accuracy was quantified using the Mean Integrated Squared Survival Error (MISSE), defined as: $\mathbb{E}\left[\int_0^{t_{\max}}\{\hat{S}(t|X) - S(t|X)\}^2 dt\right]$. MISSE evaluates the global fit of the predicted survival function $\hat{S}(t|X)$ against the true survival function S(t|X) across the entire follow-up period. MISSE was calculated for each simulated dataset. The reported MISSE is the mean across the 100 simulated datasets for each model type and drift magnitude combination. Standard errors were calculated as $SE = \frac{s}{\sqrt{100}}$, where $s$ is the SD of the MISSE values.

*Simulation Results*

In the absence of any training data from the target site, the MISSE of both the SSL and MSSL increased monotonically as the between-site heterogeneity increased (**Figure 2**). For smaller drift values (0-0.3), we saw comparable performance between both models. As the drift increased beyond 0.3, the SSL outperformed the MSSL, with the gap widening as drift increased (MISSE at drift 1.0: 0.080 vs. 0.103). Notably, the prediction errors of all models in this simulation were particularly poor as drift increased, consistent with prior work showing that, without target-site information and under substantial between-site variation, predictive performance at an unseen site is limited.[31,32]

The introduction of target site observations revealed substantial differences in model adaptability. The MSSL maintained a consistently low MISSE across all drift magnitudes and target-sample sizes (**Figure 3**; Max MISSE = 0.035). An exception was the no-drift setting, in which the MSSL was more variable than the comparators at some target sample sizes, reflecting the instability of estimating many ensemble weights from few target-site observations. In contrast, the previously best performing model (SSL) showed no ability to adapt to the target site signal, regardless of how much target site data was present (Max MISSE = 0.082). This is expected by construction since when the target-site observations were appended to the pooled training data, the small target sample is diluted by the much larger source data. The MSSL also generally outperformed or matched models developed using only the available target-site data, with the largest gains when the target sample was small and drift was large. The target-site models were highly dependent on the availability of data from the target-site to learn from. When the drift was moderate (0.5), the target-site models for $N_{\text{target}}$ = 25 and 150 had MISSEs of 0.028 and 0.008, respectively. By borrowing strength across sites, the MSSL obtained MISSEs of 0.016 and 0.006, respectively.

We performed an empirical analysis of the MSSL weight allocation to further understand its domain-adaptation properties (**Figure 4)**. Across all target sample sizes, as drift increased the MSSL reduced the weights for dissimilar source sites (1 and 2) and the increasingly noisy global pooled model, effectively learning that they were less informative for the target site. Conversely, the MSSL increasingly upweighted the similar source sites (3 and 4) as drift increased. At $N_{\text{target}}$ = 25, the target-only models, despite being fit to the target population, were highly variable and received limited weight. As $N_{\text{target}}$ increased, the weight assigned to the target-only models increased.

**Application to ADRC Data**

*Data Application Design*

We applied the MSSL to the NACC Uniform Data Set (UDS; downloaded November 5, 2025).[33] We evaluated the risk of progression from MCI to dementia across distinct ADRCs. The study cohort consisted of participants aged 50 or older with a baseline diagnosis of MCI (NACCUDSD = 3). The primary endpoint was defined as time (in days) from baseline to dementia (NACCUDSD = 4). Participants who did not progress to dementia (including who were lost to follow-up and/or died) were right censored at their final recorded visit within the UDS. There were 46 distinct study sites. After applying the inclusion criteria above, we restricted our analysis to the 10 largest ADRCs. The final dataset included five baseline predictors of cognitive decline: Age, Sex, Years of Education, Montreal Cognitive Assessment Score (MoCA), and APOE allele count (0,1,2).

We evaluated each of the 10 ADRCs as the unseen site and used the remaining 9 for training. For each target ADRC, we performed 10 random splits (Monte Carlo cross-validation) to account for sampling variability. The target ADRC was split into a 30% holdout test set and 70% for meta-learner sampling. The meta-learner was given 0, 50, 75, and 100 observations from the target ADRC to estimate the ensemble weights. For smaller target sites, where 100 > 70% of the site population, all available observations within the 70% pool were used to tune the meta-learner. The candidate learner library consists of the same models as in the simulation. The SL ensemble weights were estimated via 5-fold cross-validation using either (1) the combined training dataset (if no target site data was presented) or (2) the available target site data. IPCW was applied using site specific censoring estimates. Because the true survival function was unknown, to evaluate our model performance we used the IPCW-weighted Integrated Brier Score (IBS): $\frac{1}{t_{\max}-t_{\min}}\int_{t_{\min}}^{t_{\max}}\frac{\sum_{i=1}^{n} w_i(t)\left(\hat{S}(t|X_i)-\mathbb{I}(T_i>t)\right)^2}{\sum_{i=1}^{n} w_i(t)}\,dt$ where $w_i(t)=\frac{\mathbb{I}(T_i\le t,\Delta_i=1)}{\hat{G}(T_i^-)}+\frac{\mathbb{I}(T_i>t)}{\hat{G}(t)}$ are IPCW weights. The censoring survival function ($\hat{G}(t)$) was estimated using a marginal Kaplan-Meier model fit on the reverse event indicator $(1 - \Delta_i)$. The evaluation interval spanned the 5th and 95th percentiles of the observed event times across the source ADRCs. Across each fold, $t_{min}$ averaged to 355.3 days (Range: [352.8, 356.2]) and $t_{max}$ averaged to 2,197.8 days (Range: [2,133.0, 2,224.8]).

To assess the presence of between-site concept shift, we performed a sensitivity analysis where we fit individual Cox proportional hazards models to each of the 10 ADRCs using the same 5 covariates above. We quantified pairwise divergence between sites $i$ and $j$ as the mean absolute difference in the estimated regression coefficients across the 5 covariates: $d(i,j) = \frac{1}{5}\sum_{p=1}^{5}\left|\hat{\beta}_{i,p} - \hat{\beta}_{j,p}\right|$. For each target site $i$, we calculated its minimum distance to any other site $\left(d_{\min}(i) = \min_{j \neq i} d(i,j)\right)$ to assess if a similar source site was present in the training data. Finally, we examined the relationship between these distances and the MSSL performance to understand whether the source-target site similarity was associated with performance gains.

*Data Application Results*

The analysis cohort consisted of 1,850 subjects spanning 10 sites. Table 1 displays the baseline demographic characteristics and survival outcomes both overall and across the individual 10 ADRCs. We observed notable between-site differences in dementia conversion rates (Range: [12.3%, 43.8%]), censoring rates (Range: [57.2%, 87.7%]), and sex representation (Range (% Male): [39.3%, 66.7%]). Across six ADRCs (Sites A, B, C, E, I, and J), the MSSL had lower IBS than the SSL as more target-site data were added (**Figure 5)**. For example, at Site E the MSSL IBS decreased from 0.147 to 0.124 as the target sample increased from 0 to 100, compared with approximately 0.141 for the SSL. At Site I, the corresponding values were 0.196 to 0.184 for the MSSL versus approximately 0.197 for the SSL. In the four remaining ADRCs (Sites D, F, G, and H), the MSSL performance was equivalent or slightly worse than the SSL (**Figure 5)**. In Sites D, F, and G, the error curves are heavily intertwined between both models, suggesting little benefit from adaptation but no meaningful deterioration relative to the SSL. For Site H, the SSL performed better than the MSSL at all non-zero target sample sizes.

To better characterize these performance differences, we evaluated each target site's similarity to any source sites ($d_{\min}$). MSSL performed better on target sites with a closely matching source site, such as Site A ($d_{\min} = 0.050$) and Site G ($d_{\min} = 0.032$). Conversely, adaptation was limited on isolated target sites lacking similar training data, such as Site B ($d_{\min} = 0.208$) and Site F ($d_{\min} = 0.135$).

**Discussion**

In this paper, we developed a Multi-Site Super Learner (MSSL) to address challenges in prediction due to heterogeneity in multi-site longitudinal studies. We performed a simulation study with clusters of similar sites and showed that when a small target-site sample was available, the MSSL outperformed a SSL using pooled data under moderate to large between-site heterogeneity and generally matched or outperformed models developed using only target-site data. Without target-site data, however, the MSSL did not improve upon the SSL. The corresponding weight analysis gives both empirical and intuitive evidence for why the MSSL shows improved performance over the comparison methods. Specifically, the MSSL identifies and prioritizes source models from sites whose covariate effects resembled those of the target site. In our application to the NACC cohort, we developed predictive models for MCI to dementia. The MSSL performed comparably to or better than the SSL for 9 out of the 10 target sites, with the best performance coming in sites that had a similar site within the training data.

Our zero-shot results are consistent with evidence that, absent target information, pooled empirical risk minimization is difficult to improve upon.[35] However, pooled models assume that the target site shares the same conditional hazard distribution as the training data, which often does not hold in practice.[12] This reality motivates the continued development of domain adaptation methods, like MSSL, that identify which source data are most relevant to a new site and how best to exploit those similarities for prediction. In doing so, this approach may improve the transportability of prediction models to new centers when a small calibration sample is available.

Despite these advantages, there are several methodological limitations to consider. First, the performance gains demonstrated in this paper assume the presence of concept shift. In practice, it is difficult to know whether concept shift is present when developing a model that will extend to a new site. For practical guidance, we recommend considering the MSSL when between-site differences in covariate effects or baseline risk are plausible, for example because of differences in referral patterns, measurement protocols, or environmental exposures. Another technical limitation of this work is the assumption that a small representative calibration sample from the target site will be available. In real-world clinical settings, acquiring even a modest target sample (N = 25–100) can pose operational challenges, particularly in survival settings where observing fully characterized, labeled events (such as MCI to dementia conversion) requires sufficient longitudinal follow-up time. Furthermore, if the available target calibration data is unrepresentative due to selection bias, institutional sampling artifacts, or missing outcome data, the out-of-fold target validation matrix can become noisy and induce negative transfer. Several features of our simulation design also limit generalizability. In the simulation between-site heterogeneity was induced only through covariate effects. Baseline hazards, covariate distributions, and censoring patterns were common to all sites, and the number and size of sites were fixed. Finally, in the NACC application, the target-site test sets were relatively small, and death was not treated as a competing risk.

Future work will consider how the proposed MSSL framework situates in the broader federated learning literature. A primary direction involves adapting the MSSL's meta-weight optimization to eliminate its reliance on patient level data, thereby allowing it to be tuned to a new site without needing centralization across institutions. This could enable the construction of generalizable models across privacy-sensitive clinical networks while preserving institutional data privacy.

Acknowledgements

Funding Sources: Y.T. reports funding from P30-AG072978.

Author Contributions:

Concept and Design: H.B., H.L., and P.P.

Acquisition, analysis, or interpretation of data: all authors

Drafting of the manuscript: H.B.

Critical revision of the manuscript: All authors.

Conflicts of Interest: None to report.

Ethics and Consent to Participate declarations: not applicable

Data Sharing Statement: The data used in this study are publicly available to qualified investigators through approved applications to the National Alzheimer's Coordinating Center (NACC). Because these data are subject to data-use agreements, the authors are not permitted to redistribute the datasets or provide direct links to the analytic files used in this study. Investigators interested in accessing the data should consult the NACC website for instructions on obtaining access: https://www.naccdata.org.

NACC Data

The NACC UDS is a standardized clinical data set collected longitudinally to better characterize Alzheimer's Disease Research Center (ADRC) participants with mild Alzheimer's dementia and mild cognitive impairment compared to cognitively normal controls[26]. The data contains over 900 variables from participants, including demographics, features of symptom onset and course, personal medical history, concurrent medications, family history of dementia, and performance measures from neurological and neuropsychological examinations, that assess the major cognitive domains that are affected by AD.

The NACC database is funded by NIA/NIH Grant U24 AG072122. NACC data are contributed by the NIA-funded ADRCs: P30 AG062429 (PI James Brewer, MD, PhD), P30 AG066468 (PI Oscar Lopez, MD), P30 AG062421 (PI Bradley Hyman, MD, PhD), P30 AG066509 (PI Thomas Grabowski, MD), P30 AG066514 (PI Mary Sano, PhD), P30 AG066530 (PI Helena Chui, MD), P30 AG066507 (PI Marilyn Albert, PhD), P30 AG066444 (PI David Holtzman, MD), P30 AG066518 (PI Lisa Silbert, MD, MCR), P30 AG066512 (PI Thomas Wisniewski, MD), P30 AG066462 (PI Scott Small, MD), P30 AG072979 (PI David Wolk, MD), P30 AG072972 (PI Charles DeCarli, MD), P30 AG072976 (PI Andrew Saykin, PsyD), P30 AG072975 (PI Julie A. Schneider, MD, MS), P30 AG072978 (PI Ann McKee, MD), P30 AG072977 (PI Robert Vassar, PhD), P30 AG066519 (PI Frank LaFerla, PhD), P30 AG062677 (PI Ronald Petersen, MD, PhD), P30 AG079280 (PI Jessica Langbaum, PhD), P30 AG062422 (PI Gil Rabinovici, MD), P30 AG066511 (PI Allan Levey, MD, PhD), P30 AG072946 (PI Linda Van Eldik, PhD), P30 AG062715 (PI Sanjay Asthana, MD, FRCP), P30 AG072973 (PI Russell Swerdlow, MD), P30 AG066506 (PI Glenn Smith, PhD, ABPP), P30 AG066508 (PI Stephen Strittmatter, MD, PhD), P30 AG066515 (PI Victor Henderson, MD, MS), P30 AG072947 (PI Suzanne Craft, PhD), P30 AG072931 (PI Henry Paulson, MD, PhD), P30 AG066546 (PI Sudha Seshadri, MD), P30 AG086401 (PI Erik Roberson, MD, PhD), P30 AG086404 (PI Gary Rosenberg, MD), P20 AG068082 (PI Angela Jefferson, PhD), P30 AG072958 (PI Heather Whitson, MD), P30 AG072959 (PI James Leverenz, MD).

The NACC database is funded by NIA/NIH Grant U24 AG072122. SCAN is a multi-institutional project that was funded as a U24 grant (AG067418) by the National Institute on Aging in May 2020. Data collected by SCAN and shared by NACC are contributed by the NIA-funded ADRCs as follows:

Arizona Alzheimer's Center - P30 AG072980 (PI: Eric Reiman, MD); R01 AG069453 (PI: Eric Reiman (contact), MD); P30 AG019610 (PI: Eric Reiman, MD); and the State of Arizona which provided additional funding supporting our center; Boston University - P30 AG013846 (PI Neil Kowall MD); Cleveland ADRC - P30 AG062428 (James Leverenz, MD); Cleveland Clinic, Las Vegas – P20AG068053; Columbia - P50 AG008702 (PI Scott Small MD); Duke/UNC ADRC – P30 AG072958; Emory University - P30AG066511 (PI Levey Allan, MD, PhD); Indiana University - R01 AG19771 (PI Andrew Saykin, PsyD); P30 AG10133 (PI Andrew Saykin, PsyD); P30 AG072976 (PI Andrew Saykin, PsyD); R01 AG061788 (PI Shannon Risacher, PhD); R01 AG053993 (PI Yu-Chien Wu, MD, PhD); U01 AG057195 (PI Liana Apostolova, MD); U19 AG063911 (PI Bradley Boeve, MD); and the Indiana University Department of Radiology and Imaging Sciences; Johns Hopkins - P30 AG066507 (PI Marilyn Albert, Phd.); Mayo Clinic - P50 AG016574 (PI Ronald Petersen MD PhD); Mount Sinai - P30 AG066514 (PI Mary Sano, PhD); R01 AG054110 (PI Trey Hedden, PhD); R01 AG053509 (PI Trey Hedden, PhD); New York University - P30AG066512-01S2 (PI Thomas Wisniewski, MD); R01AG056031 (PI Ricardo

Osorio, MD); R01AG056531 (PIs Ricardo Osorio, MD; Girardin Jean-Louis, PhD); Northwestern University - P30 AG013854 (PI Robert Vassar PhD); R01 AG045571 (PI Emily Rogalski, PhD); R56 AG045571, (PI Emily Rogalski, PhD); R01 AG067781, (PI Emily Rogalski, PhD); U19 AG073153, (PI Emily Rogalski, PhD); R01 DC008552, (M.-Marsel Mesulam, MD); R01 AG077444, (PIs M.-Marsel Mesulam, MD, Emily Rogalski, PhD); R01 NS075075 (PI Emily Rogalski, PhD); R01 AG056258 (PI Emily Rogalski, PhD); Oregon Health & Science University - P30 AG066518 (PI Lisa Silbert, MD, MCR); Rush University - P30 AG010161 (PI David Bennett MD); Stanford – P30AG066515; P50 AG047366 (PI Victor Henderson MD MS); University of Alabama, Birmingham – P20; University of California, Davis - P30 AG10129 (PI Charles DeCarli, MD); P30 AG072972 (PI Charles DeCarli, MD); University of California, Irvine - P50 AG016573 (PI Frank LaFerla PhD); University of California, San Diego - P30AG062429 (PI James Brewer, MD, PhD); University of California, San Francisco - P30 AG062422 (Rabinovici, Gil D., MD); University of Kansas - P30 AG035982 (Russell Swerdlow, MD); University of Kentucky - P30 AG028283-15S1 (PIs Linda Van Eldik, PhD and Brian Gold, PhD); University of Michigan ADRC - P30AG053760 (PI Henry Paulson, MD, PhD) P30AG072931 (PI Henry Paulson, MD, PhD) Cure Alzheimer's Fund 200775 - (PI Henry Paulson, MD, PhD) U19 NS120384 (PI Charles DeCarli, MD, University of Michigan Site PI Henry Paulson, MD, PhD) R01 AG068338 (MPI Bruno Giordani, PhD, Carol Persad, PhD, Yi Murphey, PhD) S10OD026738-01 (PI Douglas Noll, PhD) R01 AG058724 (PI Benjamin Hampstead, PhD) R35 AG072262 (PI Benjamin Hampstead, PhD) W81XWH2110743 (PI Benjamin Hampstead, PhD) R01 AG073235 (PI Nancy Chiaravalloti, University of Michigan Site PI Benjamin Hampstead, PhD) 1I01RX001534 (PI Benjamin Hampstead, PhD) IRX001381 (PI Benjamin Hampstead, PhD); University of New Mexico - P20 AG068077 (Gary Rosenberg, MD); University of Pennsylvania - State of PA project 2019NF4100087335 (PI David Wolk, MD); Rooney Family Research Fund (PI David Wolk, MD); R01 AG055005 (PI David Wolk, MD); University of Pittsburgh - P50 AG005133 (PI Oscar Lopez MD); University of Southern California - P50 AG005142 (PI Helena Chui MD); University of Washington - P50 AG005136 (PI Thomas Grabowski MD); University of Wisconsin - P50 AG033514 (PI Sanjay Asthana MD FRCP); Vanderbilt University – P20 AG068082; Wake Forest - P30AG072947 (PI Suzanne Craft, PhD); Washington University, St. Louis - P01 AG03991 (PI John Morris MD); P01 AG026276 (PI John Morris MD); P20 MH071616 (PI Dan Marcus); P30 AG066444 (PI John Morris MD); P30 NS098577 (PI Dan Marcus); R01 AG021910 (PI Randy Buckner); R01 AG043434 (PI Catherine Roe); R01 EB009352 (PI Dan Marcus); UL1 TR000448 (PI Brad Evanoff); U24 RR021382 (PI Bruce Rosen); Avid Radiopharmaceuticals / Eli Lilly; Yale - P50 AG047270 (PI Stephen Strittmatter MD PhD); R01AG052560 (MPI: Christopher van Dyck, MD; Richard Carson, PhD); R01AG062276 (PI: Christopher van Dyck, MD); 1Florida - P30AG066506-03 (PI Glenn Smith, PhD); P50 AG047266 (PI Todd Golde MD PhD)

## *Tables*

Table 1: Baseline Characteristics and Outcomes across the 10 ADRCs used in the modeling.

| | **Site A (N=198)** | **Site B (N=159)** | **Site C (N=186)** | **Site D (N=128)** | **Site E (N=124)** | **Site F (N=246)** | **Site G (N=154)** | **Site H (N=177)** | **Site I (N=244)** | **Site J (N=234)** | **Overall (N=1850)** |
|---|---|---|---|---|---|---|---|---|---|---|---|
| **Clinical Outcome** | | | | | | | | | | | |
| Dementia Conversion | 36 (18.2%) | 68 (42.8%) | 71 (38.2%) | 56 (43.8%) | 29 (23.4%) | 70 (28.5%) | 65 (42.2%) | 69 (39.0%) | 30 (12.3%) | 40 (17.1%) | 534 (28.9%) |
| Censored | 162 (81.8%) | 91 (57.2%) | 115 (61.8%) | 72 (56.2%) | 95 (76.6%) | 176 (71.5%) | 89 (57.8%) | 108 (61.0%) | 214 (87.7%) | 194 (82.9%) | 1316 (71.1%) |
| **Follow-up Duration (years)** | 3.7 (1.9) | 3.0 (2.0) | 2.3 (1.7) | 2.9 (1.9) | 3.3 (1.9) | 3.3 (2.0) | 2.6 (1.8) | 3.2 (2.0) | 2.7 (1.8) | 2.8 (2.0) | 2.36 [1.23, 4.25] |
| **Age (years)** | 71.7 (7.9) | 69.8 (8.3) | 68.8 (9.4) | 73.8 (8.0) | 72.5 (8.1) | 72.5 (7.9) | 70.8 (8.3) | 79.4 (6.8) | 70.1 (7.7) | 72.7 (7.3) | 72.1 (8.4) |
| **Sex** | | | | | | | | | | | |
| Male | 85 (42.9%) | 106 (66.7%) | 95 (51.1%) | 62 (48.4%) | 80 (64.5%) | 114 (46.3%) | 72 (46.8%) | 81 (45.8%) | 108 (44.3%) | 92 (39.3%) | 895 (48.4%) |
| Female | 113 (57.1%) | 53 (33.3%) | 91 (48.9%) | 66 (51.6%) | 44 (35.5%) | 132 (53.7%) | 82 (53.2%) | 96 (54.2%) | 136 (55.7%) | 142 (60.7%) | 955 (51.6%) |
| **Education (years)** | 15.5 (2.6) | 16.4 (2.4) | 16.9 (2.9) | 16.4 (2.9) | 16.9 (2.4) | 16.0 (2.8) | 15.6 (2.6) | 16.2 (2.9) | 15.3 (3.5) | 15.9 (2.4) | 16.0 (2.8) |

Table 1: Baseline Characteristics and Outcomes across the 10 ADRCs used in the modeling.

| | Site A (N=198) | Site B (N=159) | Site C (N=186) | Site D (N=128) | Site E (N=124) | Site F (N=246) | Site G (N=154) | Site H (N=177) | Site I (N=244) | Site J (N=234) | Overall (N=1850) |
|---|---|---|---|---|---|---|---|---|---|---|---|
| **MoCA Score** | 22.5 (3.3) | 22.7 (3.4) | 23.0 (4.6) | 23.0 (3.1) | 23.7 (3.2) | 23.9 (2.9) | 21.7 (3.7) | 23.0 (2.9) | 22.3 (4.3) | 23.9 (3.2) | 23.0 (3.6) |
| **APOE ε4 Genotype** | | | | | | | | | | | |
| Carrier (≥1 ε4) | 26 (13.1%) | 17 (10.7%) | 28 (15.1%) | 16 (12.5%) | 10 (8.1%) | 32 (13.0%) | 14 (9.1%) | 25 (14.1%) | 41 (16.8%) | 33 (14.1%) | 242 (13.1%) |
| Non-carrier (0 ε4) | 172 (86.9%) | 142 (89.3%) | 158 (84.9%) | 112 (87.5%) | 114 (91.9%) | 214 (87.0%) | 140 (90.9%) | 152 (85.9%) | 203 (83.2%) | 201 (85.9%) | 1608 (86.9%) |

*Figures*

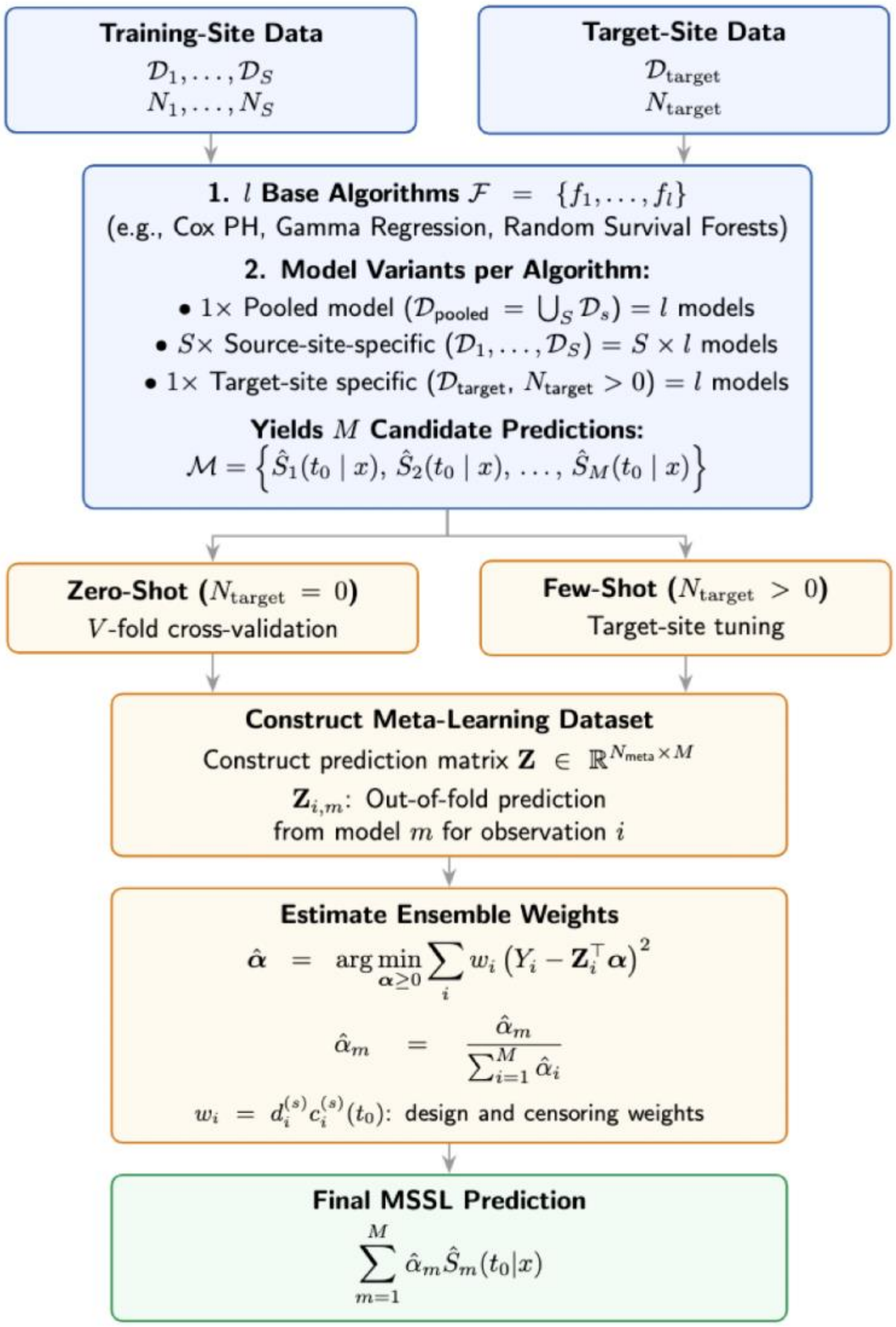

**Figure 1.** Overview of the Multi-Site Super Learner (MSSL) framework for survival prediction. Zero-shot learning occurs exclusively when no target site data is available. Few-shot adaptation and target site-specific modeling occur whenever any amount of target site data is available.

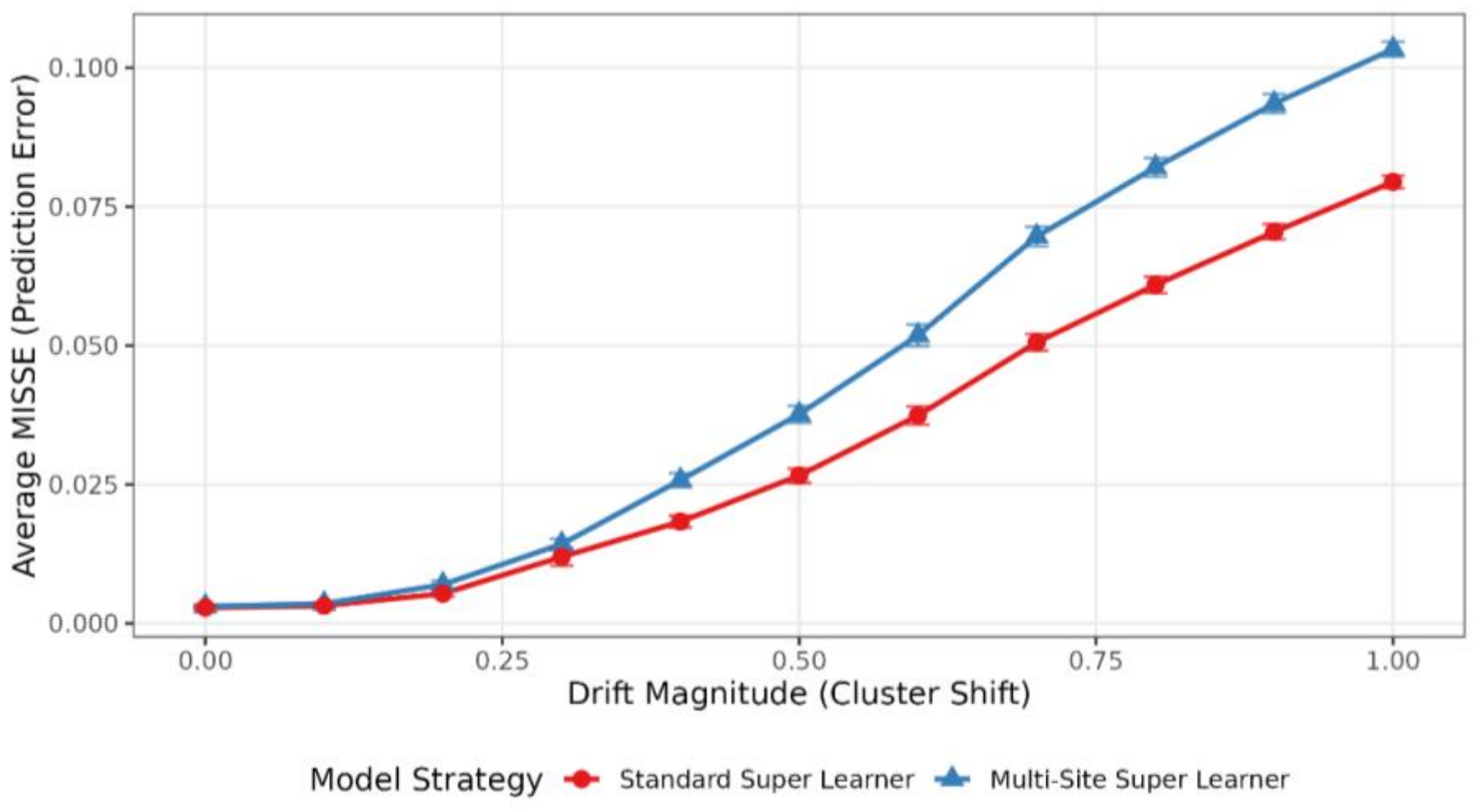

**Figure 2.** Zero-shot model performance across increasing levels of concept shift. Error bars represent $\pm 1$ SE, calculated across the 100 replicates for each model type and shift magnitude combination.

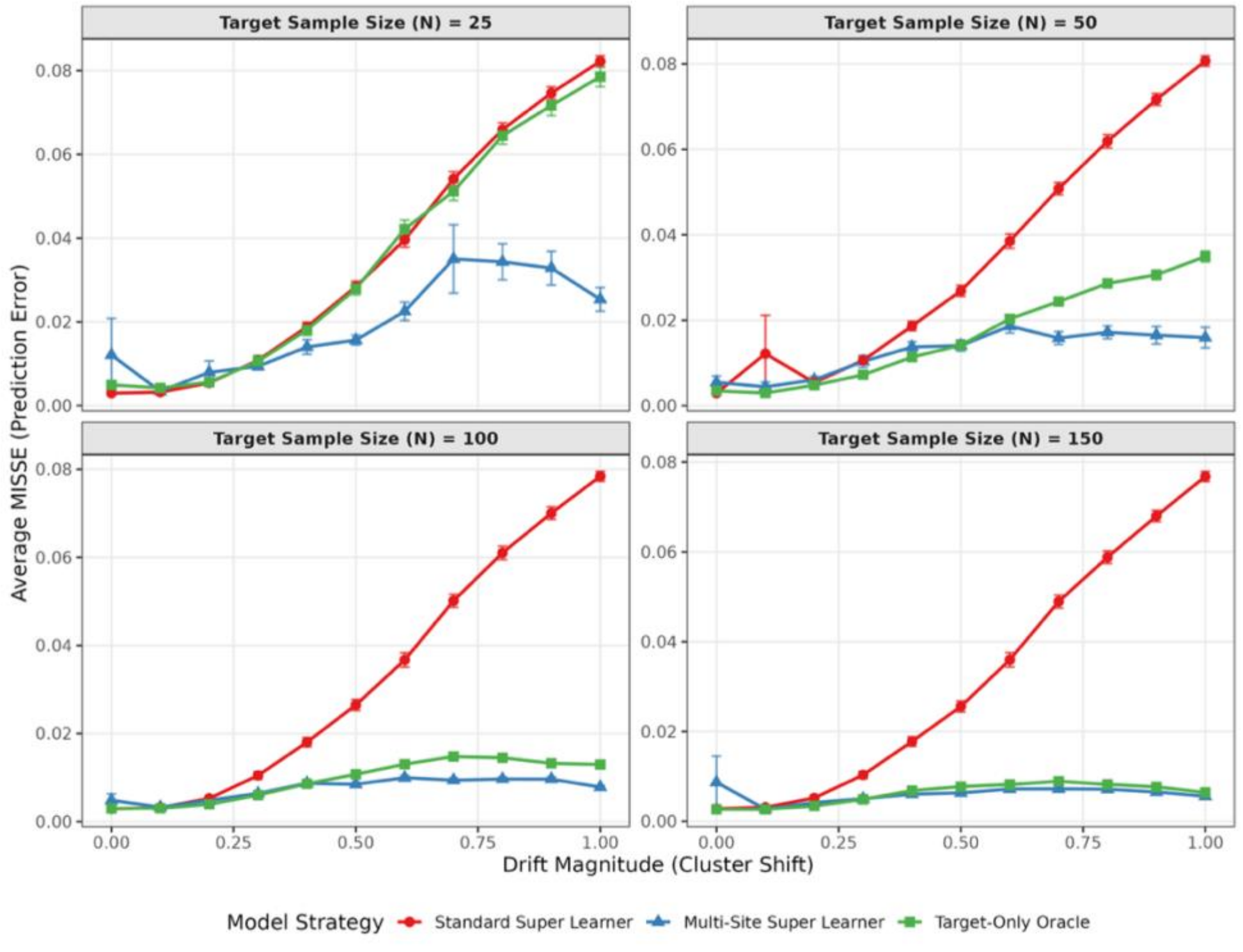

**Figure 3.** Few-shot model performance across increasing levels of concept shift. Error bars represent ±1 SE, calculated across the 100 replicates for each model type and shift magnitude combination.

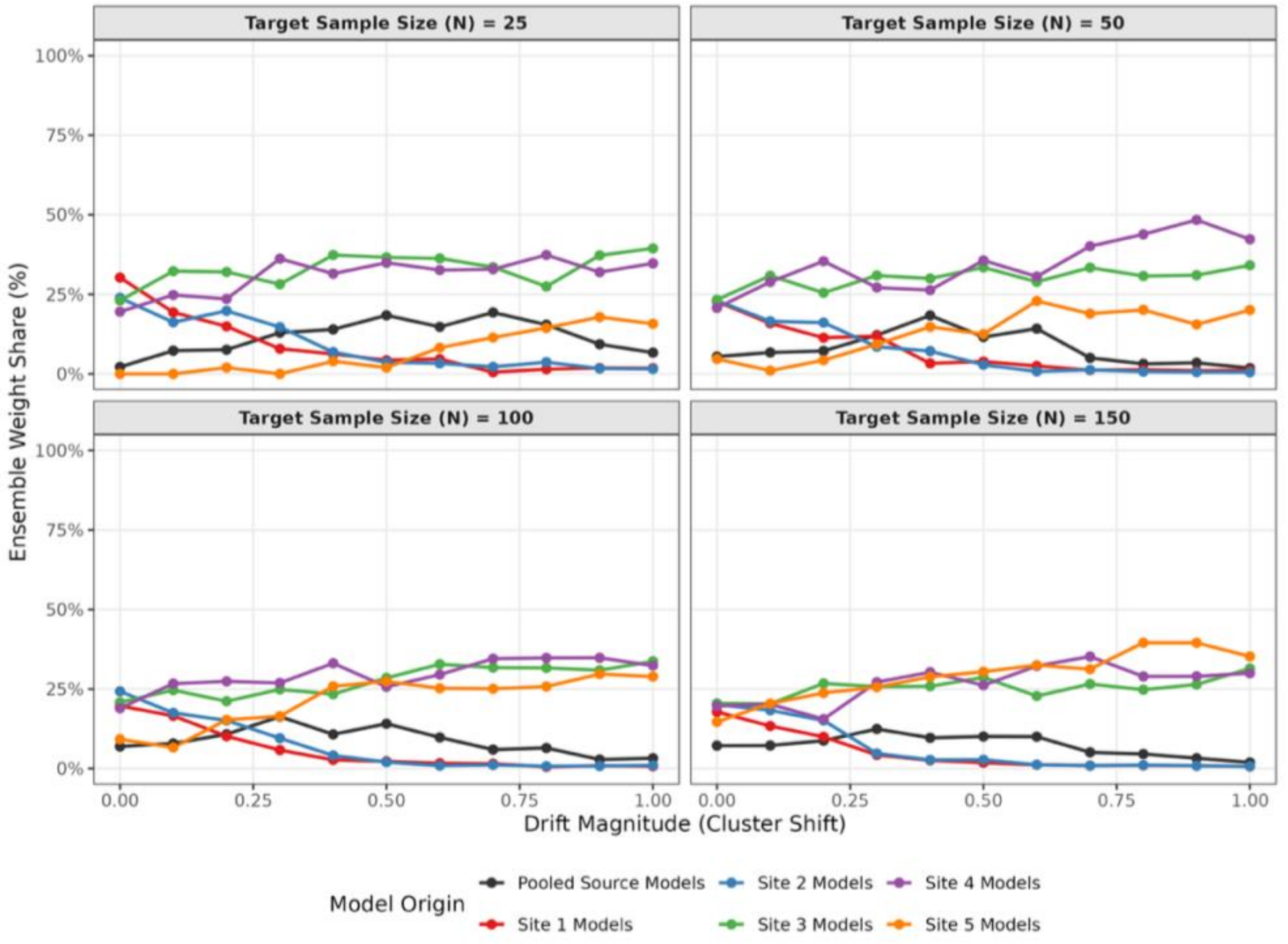

**Figure 4.** Meta-learner weight distribution across drift levels and target sample sizes. Average proportion of ensemble weight assigned to pooled source models, individual source-site models, and target-site (Site 5) models across increasing drift magnitudes (0.0 to 1.0).

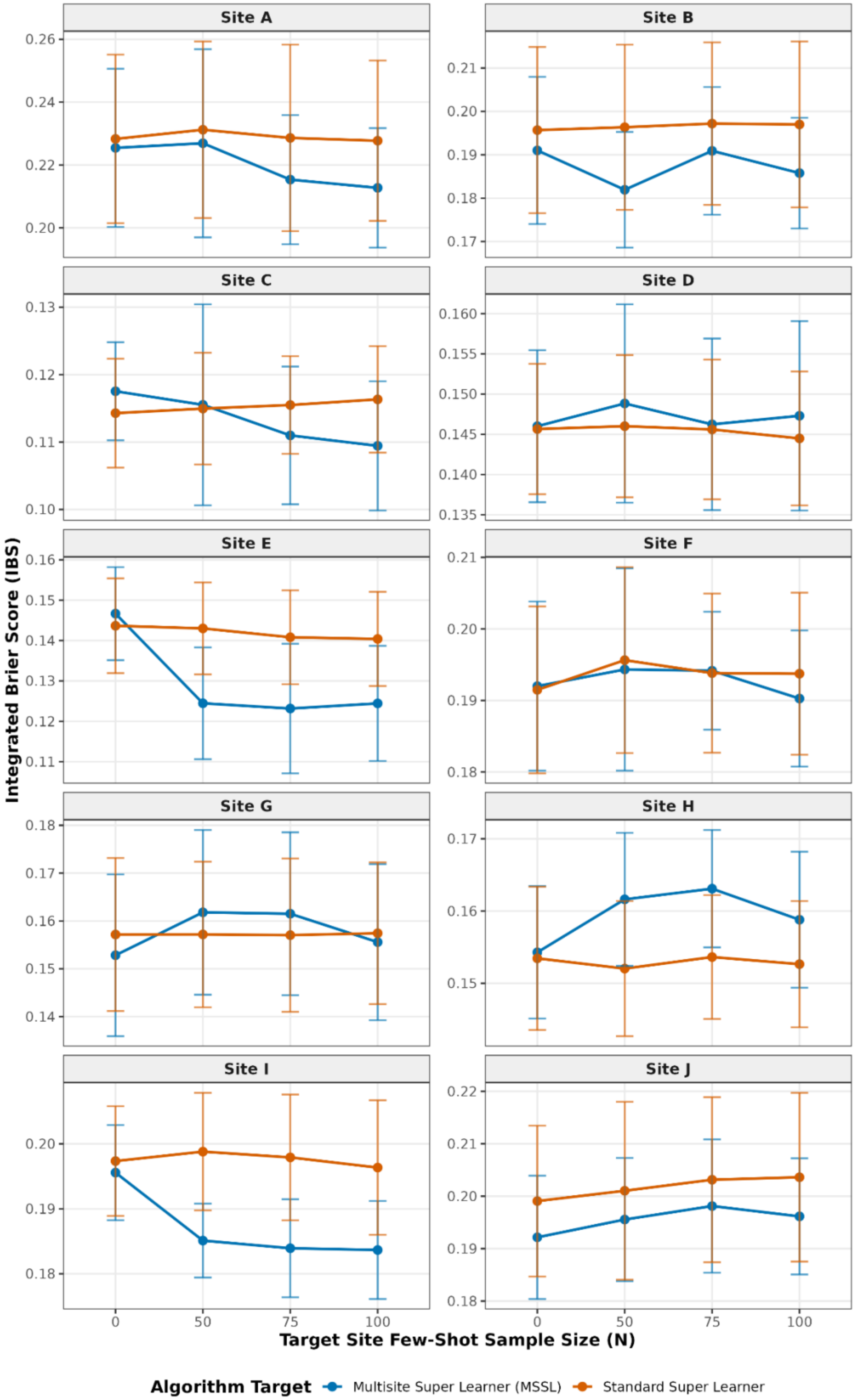
Site A
Site B
Site C
Site D
Site E
Site F
Site G
Site H
Site I
Site J
Integrated Brier Score (IBS)
Target Site Few-Shot Sample Size (N)
Algorithm Target
Multisite Super Learner (MSSL)
Standard Super Learner

20

**Figure 5.** MSSL and SSL performance on predicting MCI to dementia using NACC data. Error bars represent 95% confidence intervals across evaluation rounds.